\documentclass{PoS}
\usepackage{amsmath}
\usepackage{amssymb}
\usepackage{xfrac}
\usepackage{soul}
\pdfoutput=1
\usepackage[utf8]{inputenc}
\usepackage{epsf,amsmath,amssymb,graphicx,dcolumn}
\usepackage{caption}
\usepackage{subcaption}
\usepackage{scalefnt,ulem,pstricks}
\usepackage{booktabs,multirow,tabularx}
\usepackage{cite}
\usepackage{cleveref}
\usepackage{color}
\usepackage{rotating}
\usepackage{microtype}
\definecolor{lightblue}{rgb}{.7,.8,1}

\newcommand\nn{\nonumber}
\newcommand\f[2]{\frac{#1}{#2}} 
\newcommand{\abbrev}{\scalefont{1.}}

\def\bal#1\eal{\begin{align}#1\end{align}}

\newcommand{\as}{\alpha_{\mathrm{S}}}

\newcommand{\openloops}{{\sc OpenLoops}}
\newcommand{\munich}{{\sc Munich}}
\newcommand{\Collier}{{\sc Collier}}
\newcommand{\CutTools}{{\sc CutTools}}
\newcommand{\OneLOop}{{\sc OneLOop}}

\newcommand{\lhc}{{\abbrev LHC}}

\newcommand{\bsm}{{\abbrev BSM}}

\newcommand{\powheg}{{\abbrev POWHEG}}
\newcommand{\pythia}[1]{{\sc Pythia{#1}}}

\newcommand{\cms}{{\abbrev CMS}}

\newcommand{\zz}{\ensuremath{ZZ}}
\newcommand{\ww}{\ensuremath{W^+W^-}}

\newcommand{\vvp}{\ensuremath{VV^\prime}}
\newcommand{\system}{\ensuremath{\mathcal{F}}}
\newcommand{\angles}{\ensuremath{\Omega}}
\newcommand{\rap}{\ensuremath{y}}
\newcommand{\invm}{\ensuremath{M}}

\newcommand{\pt}{\ensuremath{p_T}}

\newcommand{\llog}{\text{\abbrev LL}}
\newcommand{\nll}{\text{\abbrev NLL}}
\newcommand{\nnll}{\text{\abbrev NNLL}}
\newcommand{\lo}{\text{\abbrev LO}}
\newcommand{\nlo}{\text{\abbrev NLO}}
\newcommand{\nnlo}{\text{\abbrev NNLO}}

\newcommand{\sm}{{\abbrev SM}}

\newcommand{\qcd}{{\abbrev QCD}}

\newcommand{\plus}{{\abbrev +}}
\newcommand{\citere}[1]{Ref.\cite{#1}}
\newcommand{\citeres}[1]{Refs.\cite{#1}}
\newcommand{\eqn}[1]{Eq.\,(\ref{#1})}

\newcommand{\fig}[1]{Fig.\,\ref{#1}}
\newcommand{\figs}[1]{Figs.\,\ref{#1}}

\newcommand{\sct}[1]{Section~\ref{#1}}

\newcommand{\dd}{{\rm d}}

\newcommand{\als}{\ensuremath{\alpha_s}}

\newcommand{\muF}{\mu_{F}}
\newcommand{\muR}{\mu_{R}}

\Crefname{figure}{Fig.}{Figs.}
\title{Transverse-momentum resummation of colorless final states at the \nnll{}\plus{}\nnlo}

\ShortTitle{Transverse momentum resummation of colorless final states at the \nnll{}\plus{}\nnlo}

\author{Marius Wiesemann\\
Physics Department, University of Z\"urich,\\
Winterthurerstrasse 190, CH-8057 Z\"urich\\
 E-mail:  \email{mariusw@physik.uzh.ch}}

\abstract{We present a general framework that allows to compute the resummed 
transverse-momentum distribution of a system of colorless  
particles. The implementation is fully differential in the 
degrees of freedom of the final-state system. 
As a first application, we consider the transverse-momentum
spectrum of \zz{} and \ww{} boson pairs produced in hadron 
collisions, where we resum the logarithmically enhanced 
contributions due to multiple soft-gluon emission at small transverse momenta
to all orders in  
perturbation theory. We exploit the most advanced perturbative information for 
the \zz{} and \ww{} production processes that is available at present by combining
next-to-next-to-leading order QCD corrections with next-to-next-to-leading 
logarithmic resummation.}

\FullConference{12th International Symposium on Radiative Corrections (Radcor 2015) and LoopFest XIV (Radiative Corrections for the LHC and Future Colliders)\\
15-19 June  2015\\
UCLA Department of Physics \& Astronomy Los Angeles, CA, USA}

\begin{document}

\section{Introduction}
\label{sec:intro}

Data collected in Run 1 and the first part of Run 2 of the Large Hadron Collider 
(\lhc) is in good agreement with the Standard Model (\sm{}) predictions so far. 
The discovery  
\cite{Aad:2012tfa,Chatrchyan:2012ufa} of a scalar resonance with a mass of 
$125$\,GeV appears to be fully consistent with the Higgs boson predicted by 
the \sm{}. This suggests that Beyond the Standard Model (\bsm{}) physics 
may appear only as small deviations from the \sm{} picture, which demands 
highly-accurate theoretical predictions.

Vector-boson pair production is an important 
class of processes at hadron colliders. They are sensitive to 
modifications of the vector-boson trilinear couplings which
arise in a large set of \bsm{} theories; they constitute an irreducible 
background to Higgs studies as well as new-physics searches. 
In particular Higgs measurements, e.g., the $H\rightarrow\ww{}$ channel, 
employ categories based on the Higgs transverse momentum 
or jet rates in order to reduce the background.\footnote{Details  
on theoretical predictions and respective uncertainties 
of these quantities can be found, e.g., in \citeres{Bagnaschi:2011tu,Harlander:2012hf,Mantler:2012bj,deFlorian:2012mx,Neumann:2014nha,Harlander:2014uea,Mantler:2015vba,Bagnaschi:2015bop,Banfi:2015pju}.}
Such analyses strongly rely on an accurate modeling of 
such observables for both signal and backgrounds.

In this article we report on a general framework to perform precision 
computations for the transverse-momentum spectrum of a system of 
colorless final-state particles implemented in the numerical code {\sc Matrix}\footnote{{\sc Matrix}
(``{\sc Munich} Automates qT subtraction and Resummation to Integrate X-sections")
is a general-purpose Monte Carlo program which combines the automated parton-level 
\nlo{} generator \munich{} \cite{munich} (``MUlti-chaNnel Integrator at Swiss~(CH) precision'')
with the $q_T$-subtraction formalism \cite{Catani:2007vq} to obtain \nnlo{} accuracy.}  \cite{matrix}.
The predictions involve next-to-next-leading order (\nnlo{}) accuracy in \qcd{} 
combined with small-\pt{} resummation at next-to-next-to-leading 
logarithmic (\nnll{}) accuracy. Besides the inclusive transverse-momentum 
spectrum, the framework allows for fiducial cuts on the colorless 
final states, owing to the fact that the 
implementation is fully-differential in the degrees of freedom of the colorless 
final-state system. This implies also the possibility to include 
off-shell effects and spin correlations when leptonic decays of any color-neutral 
boson are involved at the amplitude level. The resummation 
method is unitary \cite{Bozzi:2005wk}, so that after integration over 
\pt{} the known \nnlo{} rate is recovered.  

As a first application of the resummation framework implemented in {\sc Matrix},
the transverse-momentum distributions of on-shell \zz{} and \ww{} pairs 
at \nnll{}\plus{}\nnlo{} have been studied in \citere{Grazzini:2015wpa}, 
which is recapitulated in this report.
The \zz{} \pt{} spectrum has already been measured \cite{CMS:2014xja} 
at the \lhc{}. The resummed transverse-momentum distributions in 
\zz{} and \ww{} production have been studied before at lower perturbative and 
logarithmic accuracy in \citeres{Balazs:1998bm,Grazzini:2005vw,Frederix:2008vb,Wang:2013qua,Meade:2014fca}.

\section{Automation of transverse-momentum resummation in the {\sc Matrix} framework}
\label{sec:resum}

The general transverse-momentum 
resummation procedure was developed already in the eighties \cite{Dokshitzer:1978yd,Parisi:1979se,Curci:1979bg,Collins:1981uk,Kodaira:1981nh,Kodaira:1982az,Altarelli:1984pt,Davies:1984hs,Collins:1984kg}. For more details 
on the specific transverse-momentum resummation formalism that we implemented
in the {\sc Matrix} framework, we 
refer the reader to Refs.~\cite{Bozzi:2005wk,Bozzi:2007pn,Catani:2013tia}.

Consider a general hard-scattering process (inclusive in all parton radiation)
\begin{equation}
h_1(P_1)+h_2(P_2)\to \system(\pt,\rap,\invm)+X\,,
\end{equation}
where $\mathcal{F}$ denotes the system of an arbitrary combination 
of colorless particles produced in the collision of the two hadrons 
$h_1$ and $h_2$ with momenta $P_1$ and $P_2$, respectively.
In the center-of-mass frame 
the momentum $q=\sum_i p_i$ of the system $\system$, with the sum running over 
all particles in $\system$, is fully determined by
the invariant mass $M^2=(\sum_i p_i)^2$, the rapidity $y=\f{1}{2}\ln\f{q\cdot P_1}{q\cdot P_2}$, and the transverse-momentum \pt{}.
We shall further describe the full kinematics of each particle $i$ inside \system{} by 
the system momentum $q^\mu=\sum_i p^\mu_i$ (with $p_i^2=m_i$) and additional 
phase-space variables \angles{}. The latter do not affect the \pt{}-resummation 
procedure, but allow for a fully-differential description 
regarding the Born-level phase space, which becomes particularly relevant 
when considering leptonic final states.

With the \qcd{} factorization theorem we can write the differential cross section 
as follows:
\begin{align}
\label{dcross}
\f{\dd\sigma^{\system}}{\dd M^2\,\dd p_T^2\dd y\,\dd\angles}(y,p_T,M,\angles,s)&= \sum_{a_1,a_2}
\int_0^1 \dd x_1 \,\int_0^1 \dd x_2 \,f_{a_1/h_1}(x_1,\mu_F^2)
\,f_{a_2/h_2}(x_2,\mu_F^2)\nn\\
&\times\f{\dd{\hat \sigma}^{\system}_{a_1a_2}}{\dd M^2 \dd p_T^2\,\dd{\hat y}\,\dd\angles}({\hat y},p_T,M,\angles,{\hat s},
\as(\mu_R^2),\mu_R^2,\mu_F^2) 
\,,
\end{align}
where $f_{a/h}(x,\mu_F^2)$ ($a=q,{\bar q},g$) denotes the 
density functions of parton $a$ in hadron $h$. $\muF$ and $\muR$ are the
factorization and renormalization scales, respectively, and
$d{\hat \sigma}^{\system}_{a_1a_2}$ denotes the partonic cross section.
The rapidity ${\hat y}$ and the center-of-mass energy 
${\hat s}$ of the partonic scattering process are given by 
${\hat y}=y-\f{1}{2}\ln\f{x_1}{x_2}$ and ${\hat s}=x_1x_2 s$, where $y$ and $s$ are
their hadronic counterparts.

The transverse-momentum distribution for $\pt\gtrsim \invm$ is consistently 
described by fixed-order perturbation theory in the \qcd{} strong 
coupling constant ($\als$). When $\pt\ll\invm$ the presence of 
logarithmically-enhanced contributions $\as^n\ln^m(M^2/\pt^2)$ spoil 
the perturbative expansion in $\als$. These terms arise due to an 
incomplete cancellation of soft and collinear terms order by order 
in perturbation theory; only their all-order resummation allows 
for a physical prediction at small \pt{}.

We decompose resummation and fixed-order expansion 
at the level of the partonic cross section
\begin{equation}
\label{resplusfin}
\f{\dd{\hat \sigma}^{\system}_{a_1a_2}}{\dd M^2\,\dd p_T^2\,\dd{\hat y}\,\dd\angles}=
\f{\dd{\hat \sigma}^{\system,\rm (res.)}_{a_1a_2}}{\dd M^2\,\dd p_T^2\,\dd{\hat y}\,\dd\angles}+
\f{\dd{\hat \sigma}^{\system,\rm (fin.)}_{a_1a_2}}{\dd M^2\,\dd p_T^2\,\dd{\hat y}\,\dd\angles}\, .
\end{equation}
The first term on the r.h.s. of \eqn{resplusfin}
resums logarithmically-enhanced contributions at small \pt{}
to all orders. The second term instead contains no such 
contributions and thus remains finite as $\pt\rightarrow 0$ 
when computed in fixed-order perturbation theory.

Small-\pt{} resummation is done 
in impact-parameter ($b$) space to consistently account for both 
momentum conservation and factorization of the phase space. The resummed 
cross section is thus expressed by a Bessel transformation 
from $b$ to \pt{} space\footnote{This is strictly true only for processes induced by 
$q\bar{q}$ scattering, which are free from azimuthal correlations. In the case of 
gluon fusion this induces an additional complication at the \nnll{} accuracy \cite{Catani:2010pd}.}
\begin{equation}
\label{resum}
\f{\dd{\hat \sigma}_{a_1a_2}^{\system,(\rm res.)}}{\dd M^2\,\dd p_T^2\,\dd{\hat y}\,\dd\angles}=
\f{M^2}{\hat s} \;
\int_0^\infty \dd b \; \frac{b}{2} \;J_0(b p_T) 
\;{\cal W}^{\;\system}_{a_1a_2}(b,{\hat y},M,\angles,\hat s;\as,\mu_R^2,\mu_F^2) \;,
\end{equation}
with the $0$-order Bessel function $J_0(x)$. For simplicity we use 
Mellin moments of ${\cal W}^{\system}$. To retain the rapidity dependence, 
however, we must apply `double' $(N_1,N_2)$ Mellin moments
with respect to $z_{1,2}=e^{\pm\hat y}M/{\sqrt{\hat s}}$ as defined 
in \citere{Bozzi:2007pn}. This allows us to cast 
${\cal W}^{\system}$ in the following factorized form
\begin{align}
\label{wtilde}
&{\cal W}^{\system}_{(N_1,N_2)}(b,M,\angles;\as,\muR^2,\muF^2)
=\sigma^{\system,(0)}(\as,M,\angles)\\
&\quad\times \Bigr[ 1+\f{\as}{\pi} \,{\cal H}^{\system,(1)}_{(N_1,N_2)}(M^2/\mu^2_R,M^2/\mu^2_F,M^2/Q^2) 
\Bigr. + \Bigl.
\left(\f{\as}{\pi}\right)^2 
\,{\cal H}^{\system,(2)}_{(N_1,N_2)}(M^2/\mu^2_R,M^2/\mu^2_F,M^2/Q^2)+\dots \Bigr] \nonumber \\
&\quad\times \exp\left\{L g^{(1)}(\as L)+g^{(2)}_{(N_1,N_2)}(\as L;M^2/\mu_R^2,M^2/Q^2)\nn
+\f{\as}{\pi} g^{(3)}_{(N_1,N_2)}(\as L,M^2/\mu_R^2,M^2/Q^2)+\dots\right\}\, ,
\end{align}
where $\sigma^{\system,(0)}$ is the partonic leading-order (\lo{}) cross section. 
The coefficient functions ${\cal H}^{\system,(i)}_{(N_1,N_2)}$ of the 
$\als$ expansion 
determine all perturbative higher-order terms that behave as constants as 
$b\to\infty$, while the exponential Sudakov contains the complete dependence on 
$b$ and resums order-by-order all logarithmically-divergent $b$-dependent terms.
\eqn{wtilde} includes explicitly all terms for \nnll{} accuracy: 
$L\, g^{(1)}$ collects the \llog{} contributions, 
the function $g^{(2)}_{(N_1,N_2)}$ in combination with 
${\cal H}^{\system,(1)}_{(N_1,N_2)}$ controls the \nll{} terms, and 
$g^{(3)}_{(N_1,N_2)}$ and ${\cal H}^{\system,(2)}_{(N_1,N_2)}$ are relevant 
for \nnll{} precision. The explicit form of the resummed logarithms is given by
\begin{equation}
\label{logpar}
L=\ln\left(\f{Q^2 b^2}{b_0^2}+1\right)\, ,
\end{equation}
with $b_0=2e^{-\gamma_E}$ (and the Euler number $\gamma_E=0.5772...$).
The scale $Q$ is termed resummation scale. It parameterizes 
the ambiguities in the resummation procedure and must be chosen 
of the order of the hard scale $M$. Its variations can be exploited 
as an uncertainty estimate of yet uncalculated higher-order 
logarithmic corrections. 

Let us turn now to the finite component of the cross section (second term on the 
r.h.s of \eqn{resplusfin}), which is computed by removing 
all logarithmic terms, given by the $\als$ expansion of the resummed cross section 
in \eqn{resum}, from the customary perturbative truncation of the 
partonic cross section at a fixed-order (f.o.):
\begin{equation}
\label{resfin}
\Biggl[\f{d{\hat \sigma}^{\system,\rm (fin.)}_{a_1a_2}}{dM^2dp_T^2d{\hat y}}\Biggr]_{\rm f.o.}
=\Biggl[\f{d{\hat \sigma}^{\system}_{a_1a_2}}{dM^2dp_T^2d{\hat y}}\Biggr]_{\rm f.o.}-
\Biggl[\f{d{\hat \sigma}^{\system, \rm (res.)}_{a_1a_2}}{dM^2dp_T^2d{\hat y}}\Biggr]_{\rm f.o.}\,.
\end{equation} 
It gives the dominant contribution to the \pt{} spectrum for 
$\pt\gtrsim M$, where the fixed-order result is perfectly viable 
and any resummation effect is necessarily artificial. Indeed, 
the choice of the logarithms made in \eqn{logpar} reduces 
the impact of resummation at large \pt{}.
Moreover, for the given choice of the logarithms the argument 
of the Sudakov form factor vanishes at $b=0$, which allows us to
enforce a {\it unitarity constraint} in \eqn{resplusfin} such that 
the integration over all $\pt$ reproduces the differential 
fixed-order rate $\dd{\sigma}/{(\dd M^2\,\dd y \,\dd\angles)}$.

Finally, let us give some details on how the practical implementation and computation 
of \eqn{resplusfin} in the {\sc Matrix} framework \cite{matrix} is 
actually performed. We start from the \nlo{} calculation of $\system$+jet production 
for the fixed-order component (first term on r.h.s.) of \eqn{resfin}, 
computed with the fully-automated \nlo{} generator \munich{} \cite{munich}, 
which applies 
Catani--Seymour dipole subtraction \cite{Catani:1996vz}
and \openloops{} \cite{Cascioli:2011va} to obtain all required 
tree-level and one-loop amplitudes.\footnote{The evaluation of tensor integrals 
in the one-loop amplitudes relies on the \Collier{} library~\cite{Denner:2014gla}, 
which is numerically highly stable and based on the Denner--Dittmaier 
reduction techniques~\cite{Denner:2002ii,Denner:2005nn} and the scalar integrals of~\citere{Denner:2010tr}. For problematic phase-space points, \openloops{} provides a rescue system using the quadruple-precision implementation of the OPP method in \CutTools{} \cite{Ossola:2007ax}, involving scalar integrals from \OneLOop{} \cite{vanHameren:2010cp}.}
The \munich{} code is already combined with an automated implementation of the 
$q_T$-subtraction formalism \cite{Catani:2007vq} in the {\sc Matrix} framework
to obtain \nnlo{} accuracy, as applied in the \nnlo{} computations 
of \citeres{Grazzini:2013bna,Cascioli:2014yka,Gehrmann:2014fva,Grazzini:2015nwa}.
In fact, the finite component of \eqn{resfin} is identical in the 
$q_T$-subtraction formalism and can thus simply be taken from the \nnlo{}
implementation in the {\sc Matrix} framework. To obtain the \pt{}-resummed 
cross section in \eqn{resplusfin}, we must only replace all hard-collinear terms
(contributing at $\pt=0$) in the \nnlo{} computation by the proper 
all-order resummation formula of \eqn{resum}.

We have implemented \eqn{resum} by extending the numerical program used for 
gluon-induced Higgs production \cite{deFlorian:2012mx} such that it covers 
also the case of 
quark-initiated processes. One complication was the implementation 
of the collinear coefficients in Mellin space, which were already available 
in the code for gluon-initiated processes \cite{deFlorian:2012mx,Catani:2011kr},
while the ones relevant to quark-initiated processes, given in $x$ space in
\citere{Catani:2012qa}, we converted ourselves.

The completely general and largely automated implementation of \pt{} resummation 
in the {\sc Matrix} framework allows us to compute the resummed 
transverse-momentum spectrum for any system \system{} of colorless particles 
produced in hadron collisions up to \nnll{}\plus{}\nnlo{}, provided that the 
two-loop virtuals are available for that process. This is possible thanks to 
the fact that all the relevant resummation coefficients are known at 
sufficiently high order, and, in particular, a general relation between 
the virtual amplitudes at one and two loop and the hard function has been 
worked out up to ${\cal O}(\as^2)$ in \citere{Catani:2013tia}. In fact, 
the latter encodes all the process dependence, while the other coefficients only had 
to be determined separately for gluon- and quark-initiated processes:
The collinear coefficients can be deduced from the ones computed for
Higgs production \cite{Catani:2011kr} and Drell-Yan \cite{Catani:2012qa}; 
the universal $g^{(i)}$ functions  in \eqn{wtilde} have been expressed 
up to $i=3$ in \citere{Bozzi:2005wk} in terms of the perturbative coefficients
$A^{(1)}$, $A^{(2)}$ \cite{Catani:1988vd,Kodaira:1981nh}, $A^{(3)}$ \cite{Becher:2010tm}, ${\tilde B}_{N}^{(1)}$ \cite{Kodaira:1981nh}, ${\tilde B}_{N}^{(2)}$ \cite{Davies:1984hs,deFlorian:2000pr,deFlorian:2001zd}.

We stress again that our setup is fully differential in the momentum 
of all particles inside \system{}. Besides the possibility to study 
the \pt{} spectrum of \system{} with kinematic cuts on its constituents, 
this implies that leptonic decays of color-neutral bosons can 
be performed at the amplitude level 
including off-shell effects and spin correlations, whenever the 
two-loop helicity amplitudes are known for a process, which in turn 
allows to apply general fiducial cuts as long as they are not imposed 
on the associated jets.\footnote{This procedure required us to implement 
the recoil due to the \pt{} of the produced final-state system 
for the Born-like kinematics of the resummed component in \eqn{resum}.
We checked that our implementation is equivalent to the prescription of \citere{Catani:2015vma}.}

\section{Results: Application to \ww{} and \zz{} production}
\label{sec:results}

This Section contains the numerical results for the resummed 
transverse-momentum spectra of \vvp{} pairs with $\vvp{} \in\{\ww,\zz\}$ 
at the $\sqrt{s}=8$ TeV
\lhc{}. The only perturbative information we had to supplement to 
our framework presented in \sct{sec:resum} are the virtual amplitudes 
for the production of on-shell \ww{} and \zz{} pairs 
\cite{Cascioli:2014yka,Gehrmann:2014fva}.

Our setup uses the $G_\mu$ scheme with $G_F = 1.16639\times 10^{-5}$~GeV$^{-2}$, $m_W=80.399$ GeV and $m_Z = 91.1876$~GeV. The parton densities 
are taken from NNPDF3.0 \cite{Ball:2014uwa}.
We consider $N_f=5$ massless quarks/antiquarks for \zz{} production, while 
we employ the 4-flavor scheme for \ww{} to split off bottom-quark 
contributions in order to eliminate the contamination from $t\bar t$ 
and $Wt$ production. Our central scale choices are $\muF=\muR=\mu_0=2\,m_V$ 
for the factorization and renormalization scales, and the resummation scale 
is set to a fixed value of $Q=m_V$ as argued in \citere{Grazzini:2015wpa}.

\subsection{Inclusive transverse-momentum spectrum}
\label{sec:inclpt}

\begin{figure}[t]
\begin{center}
\begin{tabular}{cc}
\hspace*{-0.17cm}
\includegraphics[trim = 7mm -5mm 0mm 0mm, width=0.5\textwidth,]{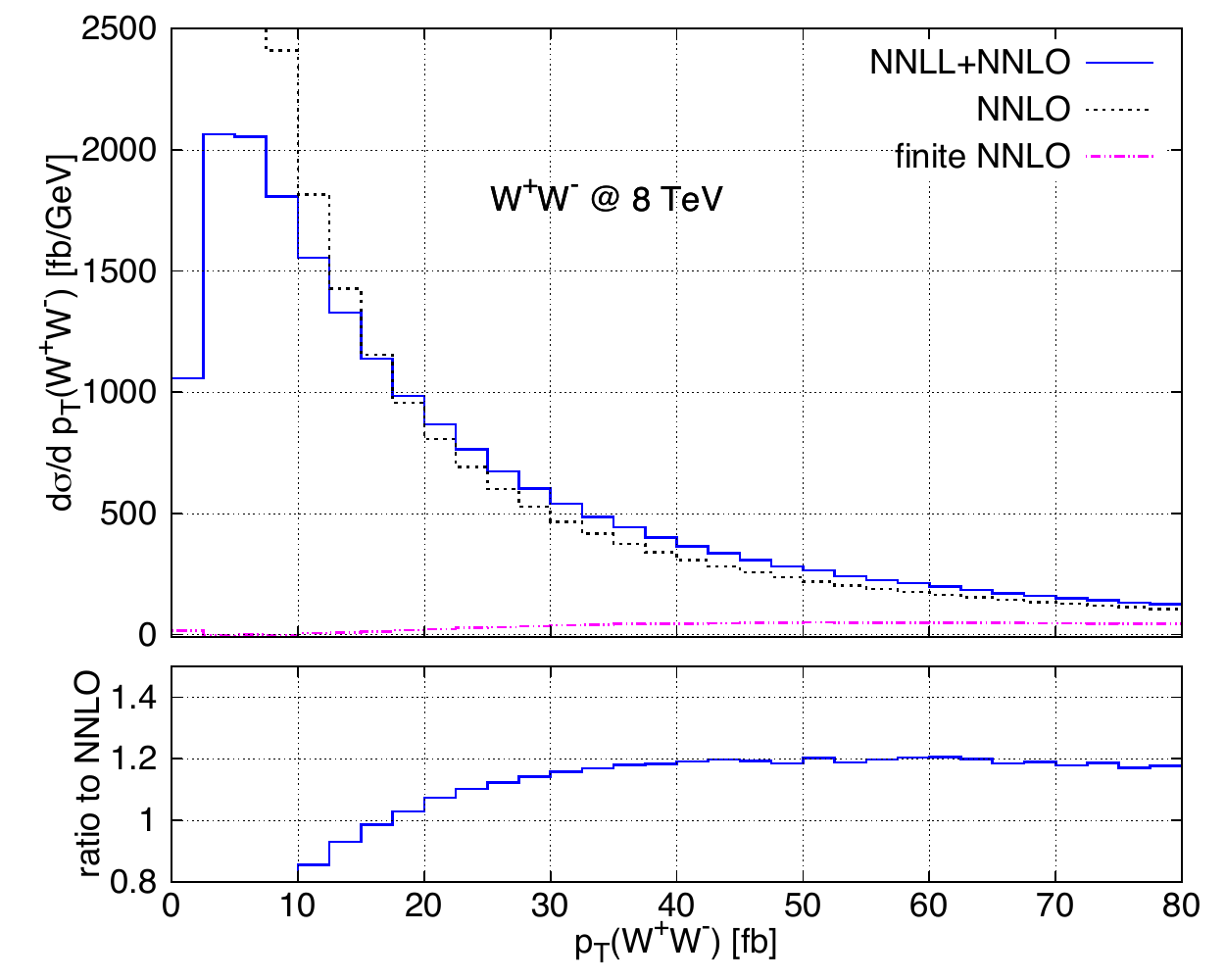} &\hspace{-0.5cm}
\includegraphics[trim = 7mm -5mm 0mm 0mm, width=0.5\textwidth]{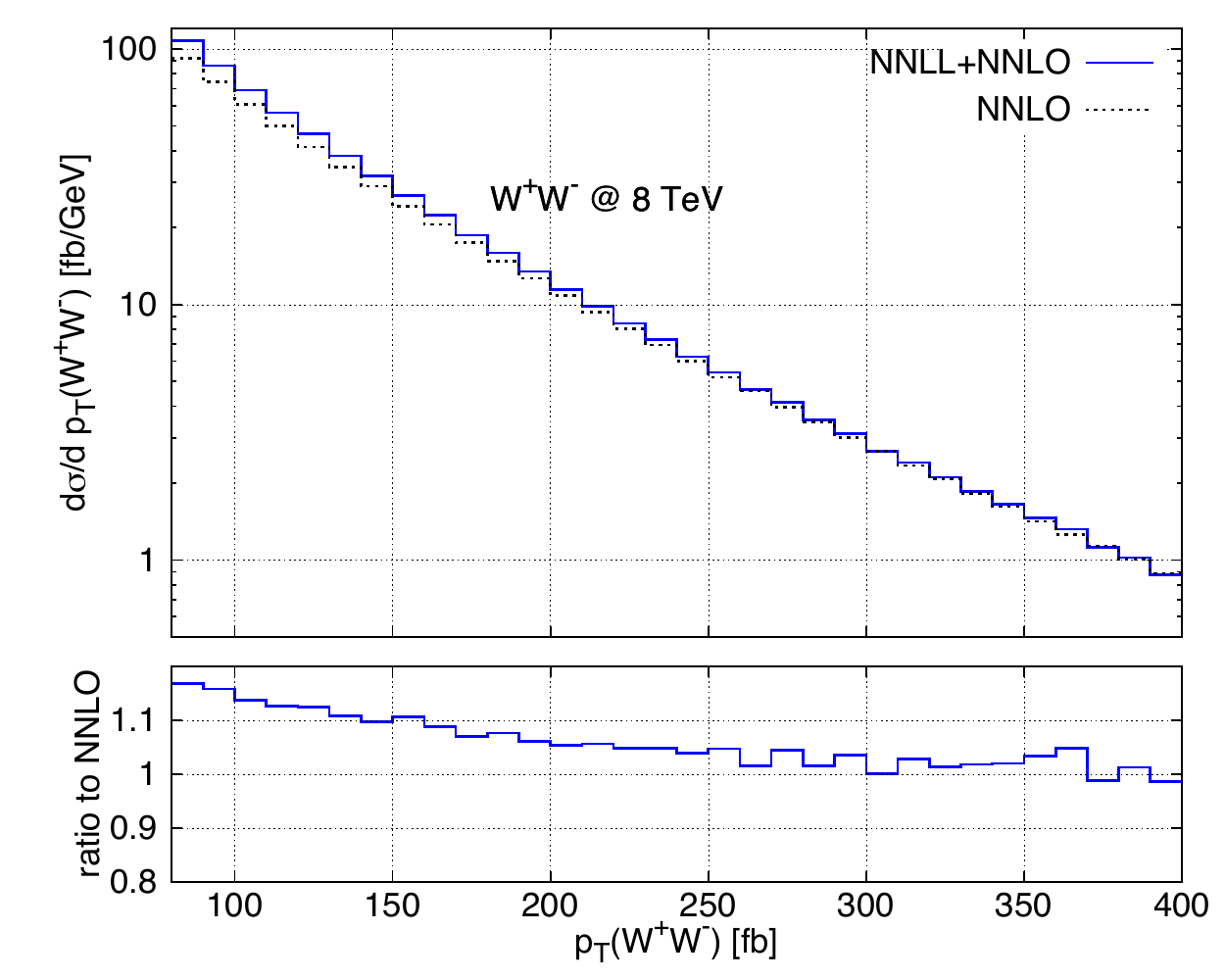} \\[-1em]
\hspace{0.6em} (a) & \hspace{-0.5cm}\hspace{1em}(b)
\end{tabular}\vspace{0.2cm}
  \parbox{.9\textwidth}{%
      \caption[]{\label{fig:matchingNNLO}{\sloppy 
      \nnll{}\plus{}\nnlo{} \pt{} spectrum (blue, solid)
       of the \ww{} pair at (a) small and (b) large \pt{} is compared to 
       \nnlo{} (black, dotted) and the finite component of 
       \eqn{resplusfin} (magenta, dash-double dotted). 
       The lower insets show the \nnll{}\plus{}\nnlo{} to \nnlo{} ratio.}}}
\end{center}
\end{figure}

\begin{figure}[t]
\begin{center}
\hspace*{-0.15cm}
\begin{tabular}{cc}
\includegraphics[trim = 7mm -5mm 0mm 0mm, width=0.5\textwidth,]{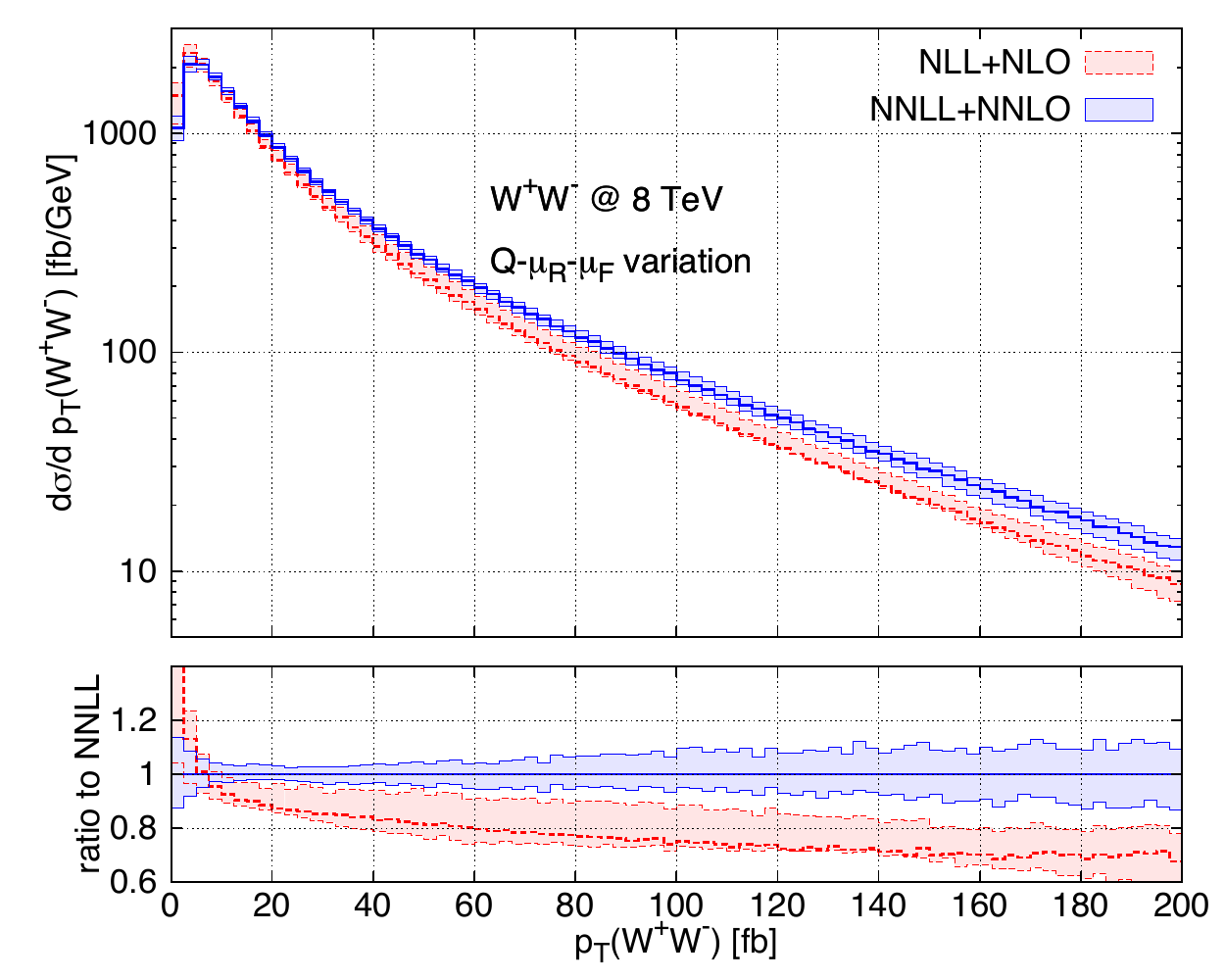} &\hspace{-0.5cm}
\includegraphics[trim = 7mm -5mm 0mm 0mm,width=0.5\textwidth]{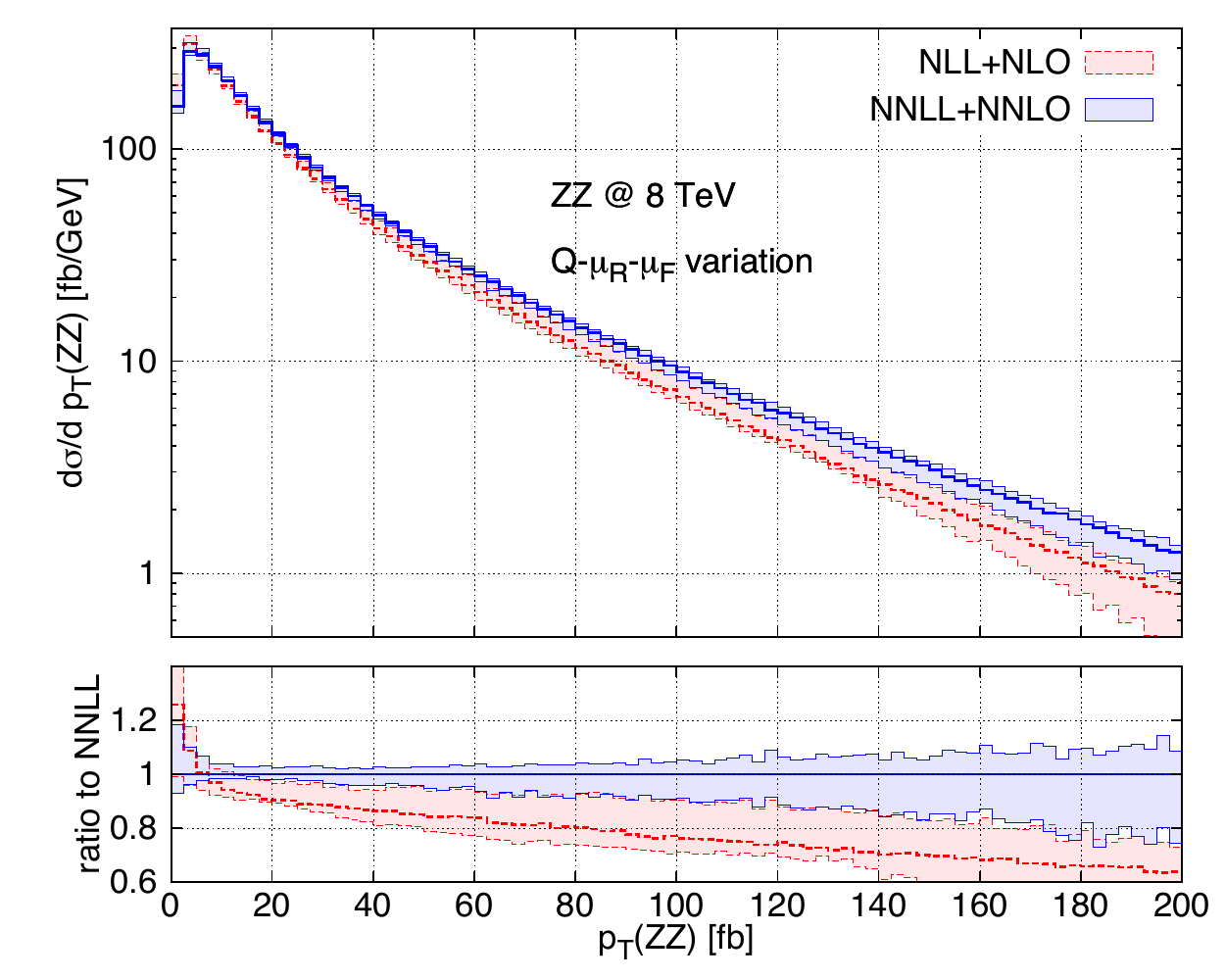} \\[-1em]
\hspace{0.6em} (a) & \hspace{-0.5cm}\hspace{1em}(b)
\end{tabular}\vspace{0.2cm}
  \parbox{.9\textwidth}{%
      \caption[]{\label{fig:bestpredictionww}{
            \sloppy  \pt{} spectrum of (a) the \ww{} pair and (b) the \zz{} pair
            at \nll{}\plus{}\nlo{} (red, dashed) and \nnll{}\plus{}\nnlo{}
          (blue, solid);
  thick lines: central prediction; bands: $\muF{}$, $\muR{}$ and $Q$ uncertainties 
 computed as described in the text; thin lines: borders of bands. The lower insets show the ratio to \nnll{}\plus{}\nnlo{}.
}}}
\end{center}
\end{figure}

We start by comparing the resummed \nnll{}\plus{}\nnlo{} prediction (blue, solid) 
for the inclusive \ww{} transverse-momentum distribution to the \nnlo{} 
result (black, dotted) in \fig{fig:matchingNNLO}.\footnote{The general 
considerations apply also to \zz{} production so that it is sufficient to
focus on \ww{} \pt{} spectra at first.}
 As expected, the fixed-order curve
diverges at small \pt{} and provides no physical prediction in that region.
The resummed result, on the other hand, has a well behaved spectrum 
down to vanishing transverse momenta. Its spectrum features a 
kinematical peak at $p_T\sim 5$ GeV. At low transverse momenta 
($\pt \le 80$\,GeV) in \fig{fig:matchingNNLO}\,(a), we also study the 
impact of the finite component (cf. \eqn{resplusfin}) to the resummed 
distribution (magenta, dash-double dotted), which contributes below 
$1\%$ in the peak region and $\sim19\%$ at $p_T=50$ GeV.

Looking at the ratio of fixed-order and resummed predictions 
at large transverse momenta ($80$\,GeV$\le\pt\le 400$ GeV) 
in the lower inset of \fig{fig:matchingNNLO}\,(b), we see that the 
\nnll{}\plus{}\nnlo{} distribution smoothly merges into the \nnlo{} result.
We checked that this behaviour is indeed preserved up to very large 
transverse momenta, which, in fact, renders a hard switching \cite{Harlander:2014hya} to the 
fixed-order result feasible. Therefore, the \nnll{}\plus{}\nnlo{} computation provides 
a uniform prediction which consistently combines low- and high-\pt{} results.

We now turn to our best prediction for \ww{} and \zz{} transverse-momentum 
spectra including scale uncertainties that are shown 
in \figs{fig:bestpredictionww}\,(a) 
and (b), respectively. We compare the resummed \nnll{}\plus{}\nnlo{} 
result (blue, solid) to \nll{}\plus{}\nlo{} (red, dashed). The uncertainty bands
reflect the combined uncertainty from independent $\muF$, $\muR$ and $Q$ variations
in the ranges $m_V\leq \{\muF,\muR\} \leq 4m_V$ and $m_V/2 \leq Q \leq 2m_V$, 
while constraining $0.5\leq \muF/\muR\leq 2$ and $0.5\leq Q/\muR\leq 2$. 
By and large we find a consistent reduction of the residual uncertainties:
For \ww{} the uncertainty at \nnll{}\plus{}\nnlo{} (\nll{}\plus{}\nlo{}) 
amounts to about $\pm 8\%$ ($\pm 12\%$) at the peak, $\pm 3\%$ ($\pm 5\%$) at $p_T=20$ GeV and $\pm 10\%$ ($\pm 15\%$) at $p_T=200$ GeV; in case of \zz{}
the pattern of the uncertainties in the small- and intermediate-\pt{} region
is very similar; only at large transverse momenta they are larger 
than for \ww{} reaching up to about $\pm 17\%$ at 
\nnll{}\plus{}\nnlo{} for $p_T=200$ GeV.

The behaviour in the large-\pt{} region is driven by the fixed-order distribution. 
Let us recall that for \ww{} the \nnlo{} corrections \cite{Dittmaier:2007th,Campbell:2007ev,Dittmaier:2009un} for $\pt\lesssim 200$\,GeV
with respect to \nlo{} are quite large ($\sim 40\%)$ and that at 
least the \nlo{} scale 
variations underestimate considerably the theoretical 
uncertainty, given the fact the \nlo{} and \nnlo{} bands do not overlap 
in that region \cite{Grazzini:2015wpa}. Therefore, it is not surprising that 
we find non-overlapping and hardly-overlapping bands at large transverse 
momenta for \ww{} and \zz{}, respectively.

\subsection{Rapidity dependence of the transverse-momentum spectrum}
\label{sec:rap}

\begin{figure}[h]
\begin{center}
\begin{tabular}{cc}
\includegraphics[trim = 7mm -5mm 0mm 0mm, width=0.495\textwidth]{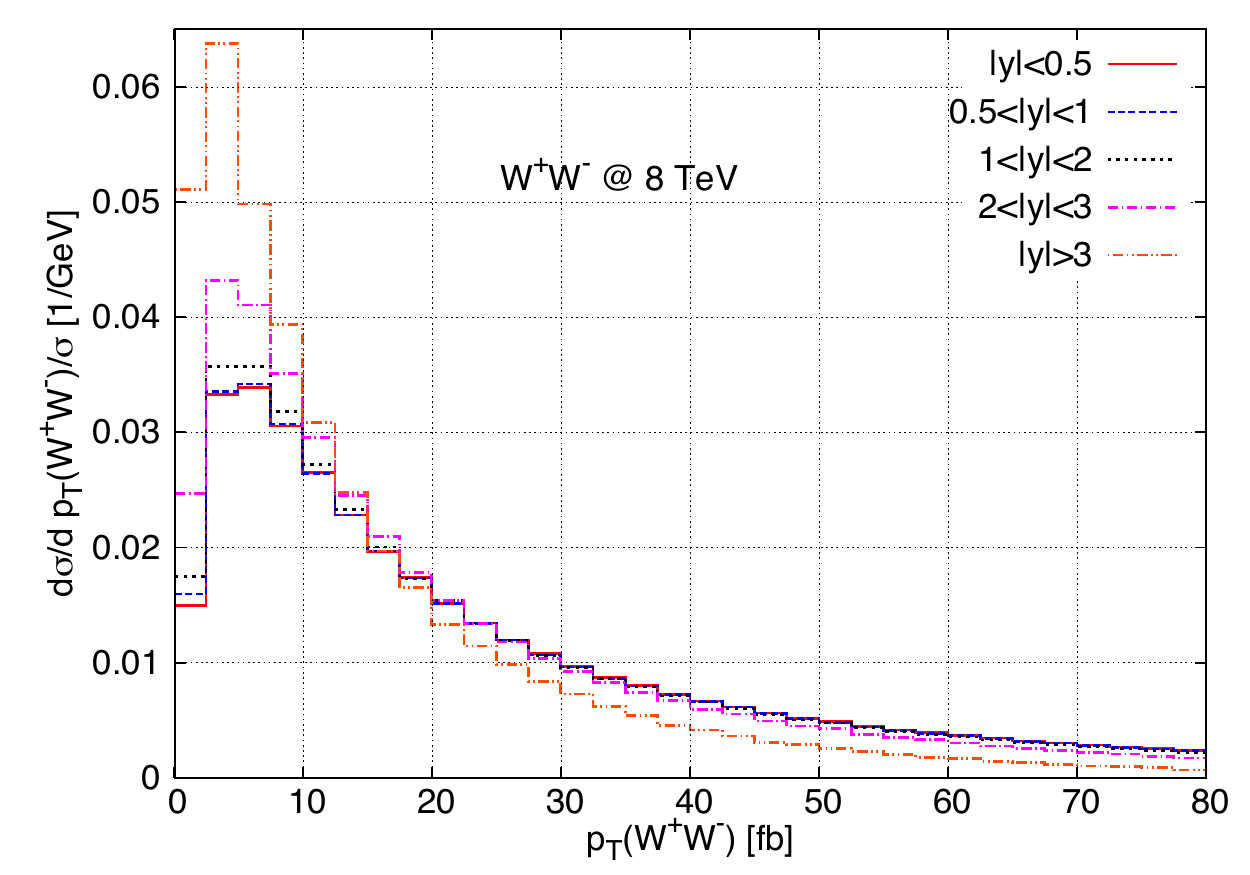} & \hspace{-0.5cm}
\includegraphics[trim = 7mm -5mm 0mm 0mm, width=0.495\textwidth]{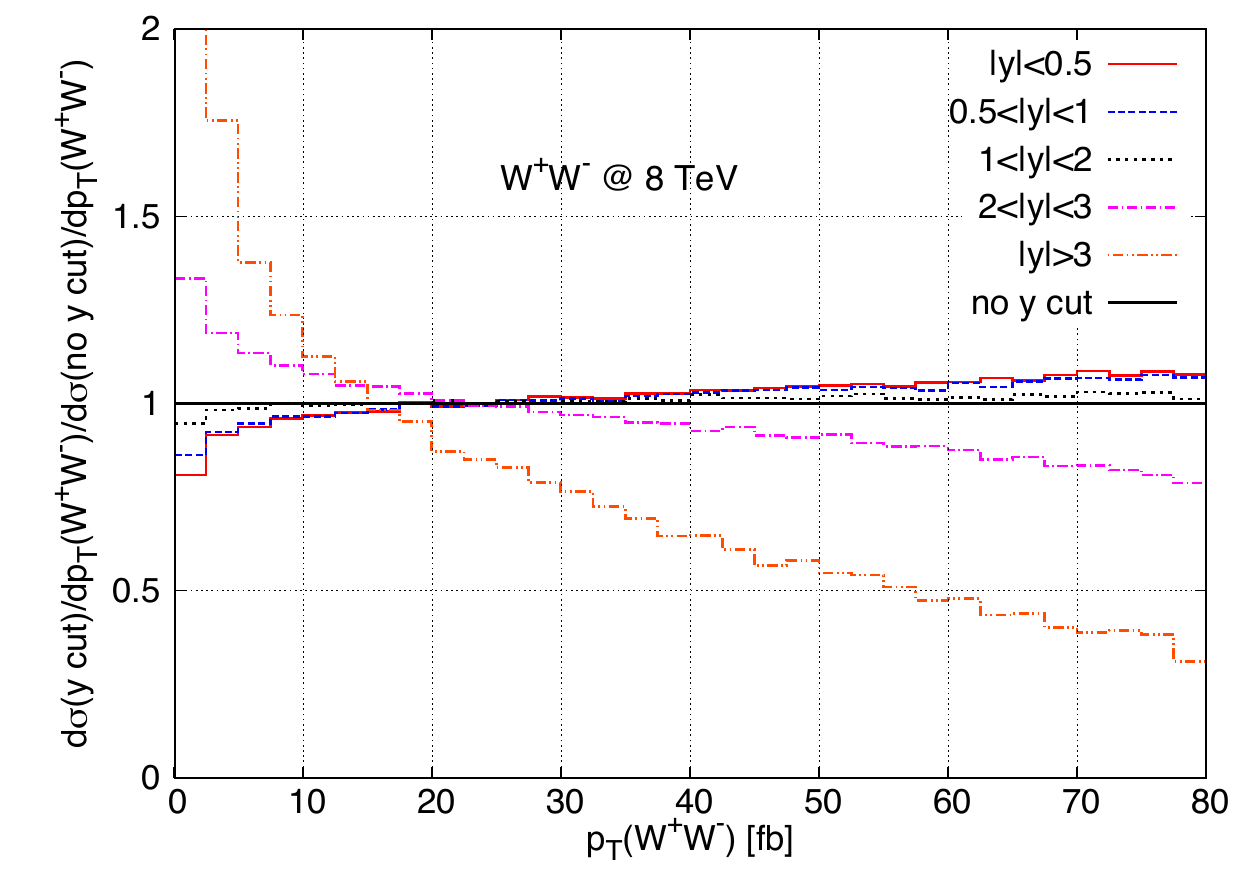} \\[-1em]
\hspace{0.6em} (a) & \hspace{-0.5cm}\hspace{1em}(b)
\end{tabular}\vspace{0.2cm}
  \parbox{.9\textwidth}{%
      \caption[]{\label{fig:rap}{\sloppy  (a) \ww{} transverse-momentum shapes 
          at \nnll{}\plus{}\nnlo{} with cuts on the rapidity of the \ww{} pair:  
          $|y|<0.5$ (red, solid), $0.5<|y|<1$ (blue, dashed), $1<|y|<2$ 
          (black, dotted), $2<|y|<3$ (magenta, dash-dotted) and $3<|y|$ 
          (orange, double-dash dotted); (b) shape-ratio with 
          respect to the inclusive spectrum.
}}}
\end{center}
\end{figure}

As described in \sct{sec:resum} our implementation of the general 
resummation formalism is fully differential in the \vvp{} phase space, 
i.e., it allows for arbitrary cuts on the kinematics of the 
\vvp{} pair (and even on any of its decay products, once we include 
the leptonic \vvp{} decays by applying the helicity amplitudes 
of \citeres{Caola:2015ila,vonManteuffel:2015msa}). A natural 
double-differential observable concerns the \vvp{} \pt{} distribution 
with an additional cut on the rapidity of the vector-boson pair.

\fig{fig:rap}\,(a) shows the shape, i.e., normalized such that its integral yields one,
of \nnll{}\plus{}\nnlo{} \pt{} distributions of \ww{} pairs in various rapidity ranges: 
$|y|<0.5$ (red, solid), $0.5<|y|<1$ (blue, dashed), $1<|y|<2$ (black, dotted), 
$2<|y|<3$ (magenta, dash-dotted) and $3<|y|$ (orange, dash-double dotted). 
In \fig{fig:rap}\,(b) these curves are normalized to the shape of the 
inclusive \pt{} distribution. The general observations are the following:
\begin{itemize}
\item The \pt{} shapes are hardest at central rapidities and become softer as the 
rapidity increases.
\item In the central region ($|y|<2$) the shape of the \ww{} transverse momentum 
spectrum is rather insensitive to the specific rapidity value. Indeed, the curves 
become only slightly harder than in the inclusive case.
\item In the forward rapidity region the curves feature a significant distortion 
towards a softer spectrum with respect to the inclusive shape; with deviations of 
more than a factor of two in the shape-ratio. These effects, however, are strongly 
phase-space suppressed. 
\end{itemize}

The observed pattern can be understood in the following way: rapidity 
and transverse momentum are two not completely independent phase-space 
variables. Indeed, they affect their mutual upper integration bounds. At higher 
rapidities the kinematically allowed range of transverse momenta is reduced:
this squeezes the \pt{} spectrum which consequently 
becomes softer. This effect has been observed also in previous studies
in the case of Higgs boson production \cite{Bozzi:2007pn}.

\subsection{\pt{}-veto efficiencies for \ww{} production}
\label{sec:pteff}

\begin{figure}[h]
\begin{center}
\includegraphics[trim = 10mm -5mm 0mm 0mm, height=.35\textheight]{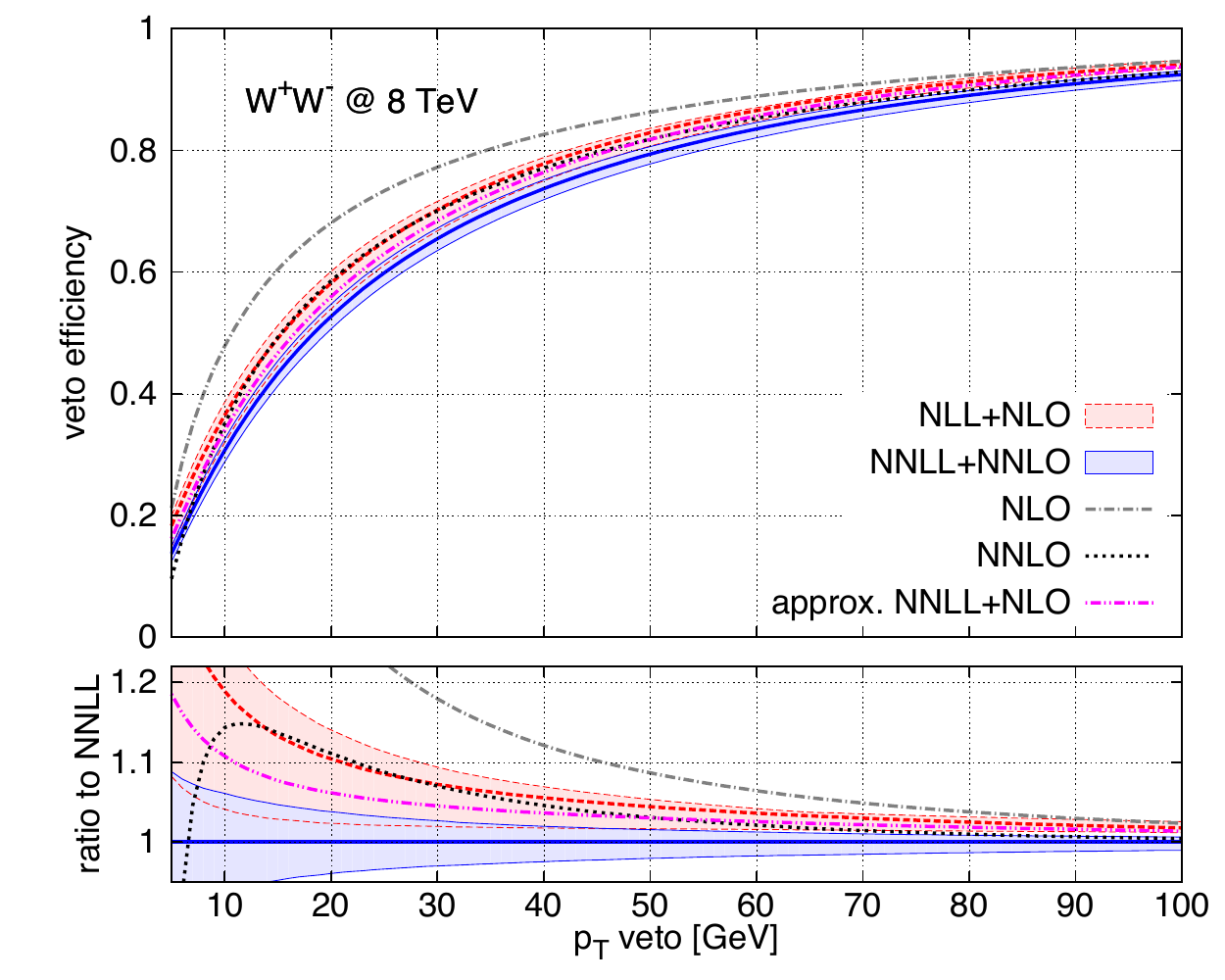}
    \parbox{.9\textwidth}{%
      \caption[]{\label{fig:veto}{\sloppy \pt{}-veto efficiency of the \ww{} pair 
      at various orders: \nll{}\plus{}\nlo{} (red, dashed), 
      \nnll{}\plus{}\nnlo{} (blue, solid), \nlo{} (grey, dash-dotted), \nnlo{} (black, dotted),
       approximate \nnll{}\plus{}\nlo{} (magenta, dash-double dotted); thick lines: central prediction; bands: uncertainty
          due to combined scale variations; thin lines: borders of bands.}  }}
\end{center}
\end{figure}

In this section we study efficiencies of the transverse-momentum of the
\ww{} pair, defined as
\bal
\epsilon(\pt^{\text{veto}}) = \sigma(\pt<\pt^{\text{veto}}{})/\sigma_\text{tot}\,,
\eal
at various orders in resummed and fixed-order perturbation theory. \fig{fig:veto} 
shows predictions for $\epsilon(\pt^{\text{veto}})$ 
at the \nnll{}\plus{}\nnlo{} (blue, solid), 
approximate \nnll{}\plus{}\nlo{} (magenta, dash-double dotted), 
\nll{}\plus{}\nlo{} (red, dashed), \nnlo{} (black, dotted) 
and \nlo{} (grey, dash-dotted) as a function of $\pt^{\text{veto}}$.
In the lower inset the results are normalized to the reference \nnll{}\plus{}\nnlo{}
prediction. Approximate \nnll{}\plus{}\nlo{} denotes 
\nll{}\plus{}\nlo{}, but adding the $g^{(3)}$ function in the Sudakov exponent 
in \eqn{wtilde}, and corresponds to the approximation considered in 
\citeres{Grazzini:2005vw,Meade:2014fca}. The uncertainty bands involve the
independent variations of $\muF$, $\muR$ and, where applicable, 
$Q$, as described in \sct{sec:inclpt}.

The general observation is that both resummation and perturbative higher-order 
effects yield a sizable reduction of the \pt{}-veto efficiency and 
therefore are vital for a precise theoretical prediction of that quantity. 
Indeed, the approximated \nnll{}\plus{}\nlo{} result gives some improvement over 
the \nll{}\plus{}\nlo{} one, but is still roughly $5\%$ above the reference 
prediction at \nnll{}\plus{}\nnlo{} for $\pt^{\text{veto}}\sim 25-30$ GeV.
This suggests that the jet-veto efficiency obtained from the 
reweighting of \powheg{} \cite{Nason:2004rx} plus 
\pythia{6}{} \cite{Sjostrand:2006za} with the approximate \nnll{}\plus{}\nlo{} 
result of the inclusive \ww{} \pt{} spectrum in \citere{Meade:2014fca}, 
which was used in the \ww{} measurement by \cms{} \cite{CMS:2015uda}, might 
decrease when using the full \nnll{}\plus{}\nnlo{} prediction.

\subsection{Comparison to data of the \zz{} spectrum}
\label{sec:zzcomparison}

\begin{figure}[h]
\begin{center}
\begin{tabular}{cc}
\includegraphics[trim = 7mm -5mm 0mm 0mm, width=0.46\textwidth]{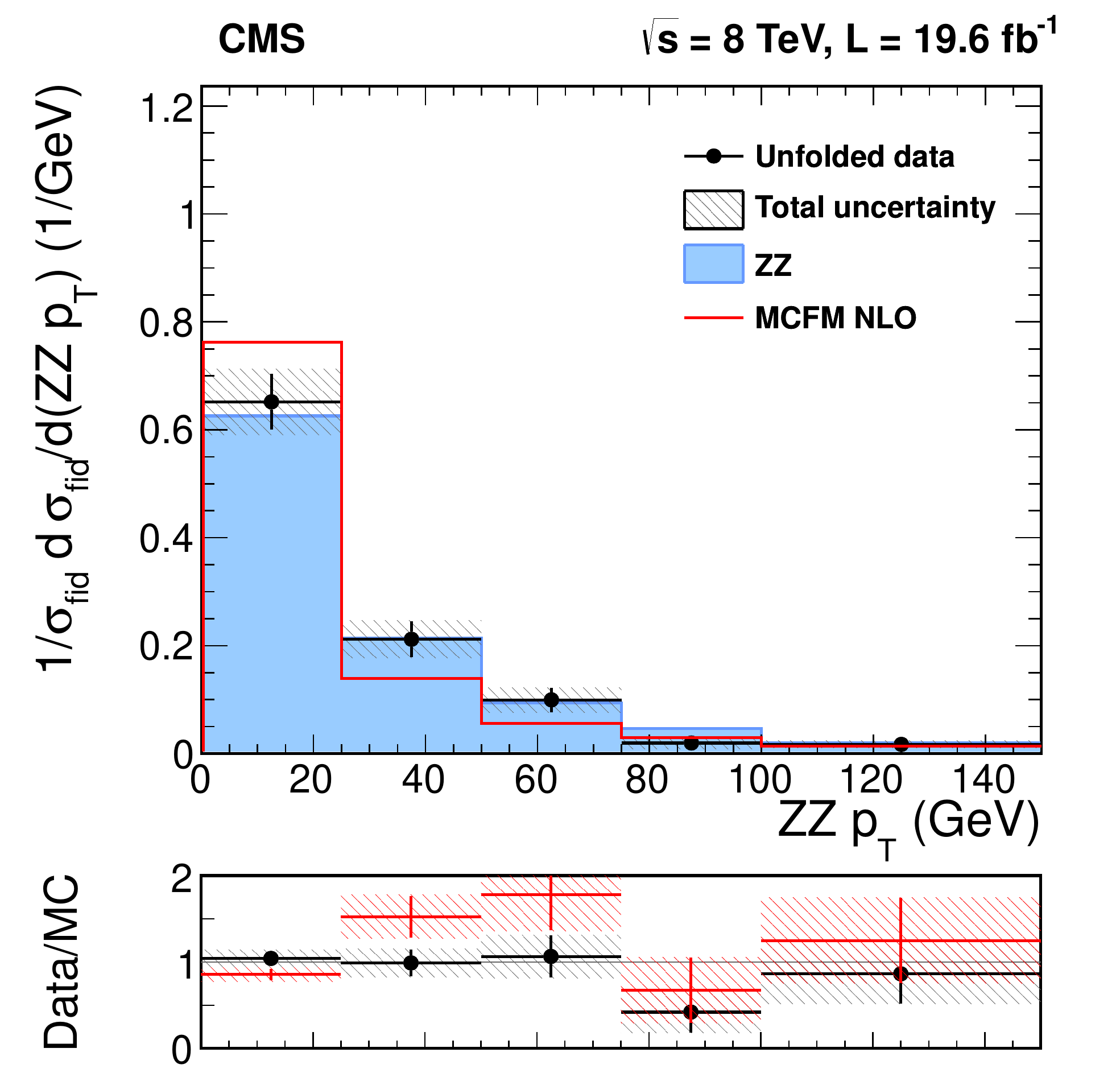} & 
\includegraphics[trim = 7mm -5mm 0mm 0mm, width=0.51\textwidth]{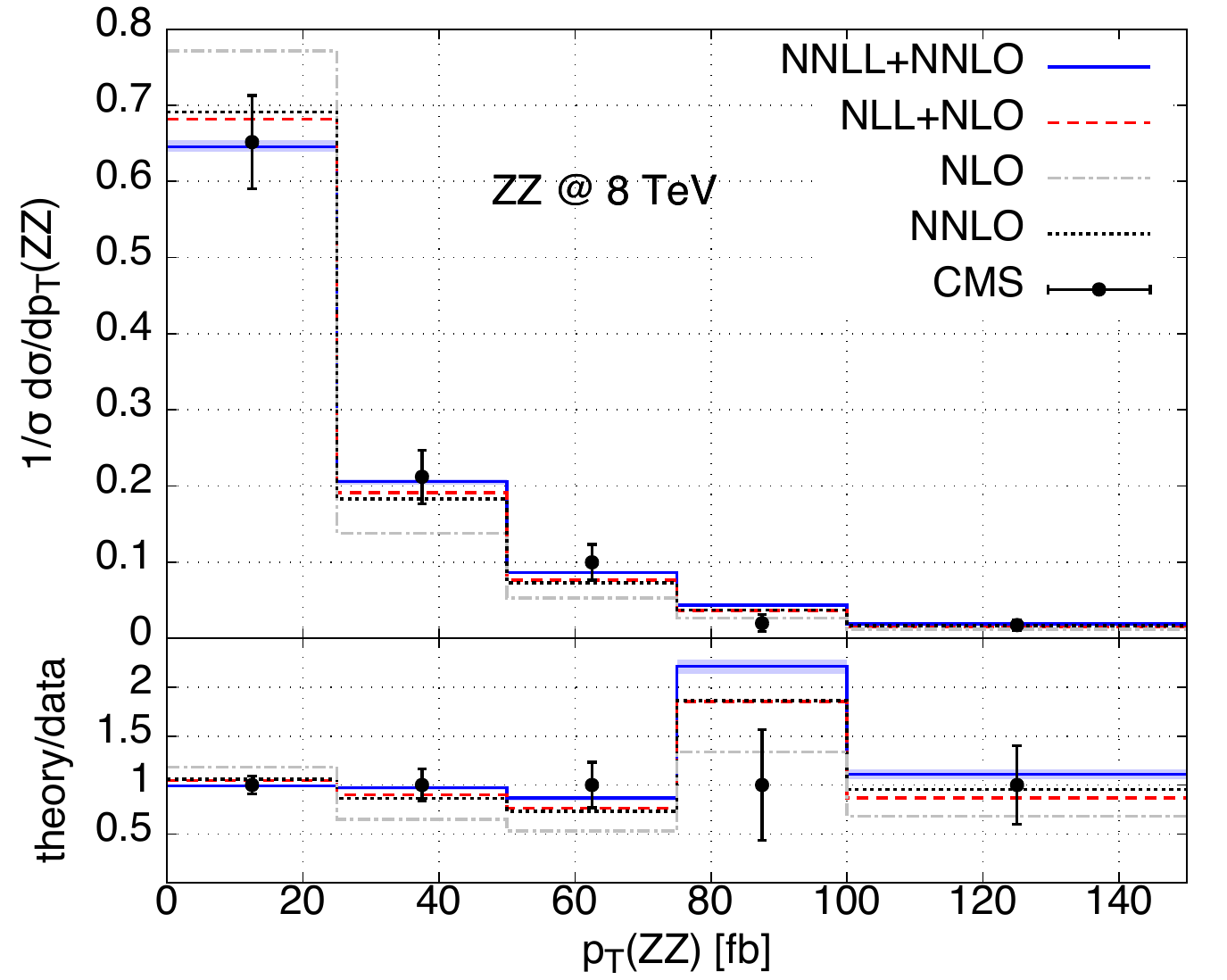} \\[-1em]
\hspace{0.6em} (a) & \hspace{1em}(b)
\end{tabular}\vspace{0.2cm}
  \parbox{.9\textwidth}{%
      \caption[]{\label{fig:data}{\sloppy  (a) Experimental measurement 
      of the \zz{} \pt{} shape in the fiducial region from \citere{CMS:2014xja}
	  and (b) comparison of the data with various predictions at higher orders.
}}}
\end{center}
\end{figure}

In \fig{fig:data} we compare the experimental measurement of the 
\zz{} \pt{} distribution by \cms{} presented in \citere{CMS:2014xja}
to predictions at various orders in resummed and fixed-order perturbation 
theory. Let us stress that the comparison is done at the level of shapes, 
more precisely the bins add up to one, and that the comparison is not 
completely consistent, since the experimental \pt{} shape is measured in the 
fiducial volume, while our predictions are for the fully-inclusive spectrum.
Fiducial cuts are not expected to change the picture dramatically though.

That being said, we observe a remarkable agreement between our best 
\nnll{}\plus{}\nnlo{} prediction (blue, solid curve) and the data 
points (black dots), except for the single bin ($75$\,GeV$\le\pt\le 100$\,GeV) 
where the experimental uncertainties are largest. Even in this bin 
the deviation is still below the two sigma level though. We note that also the 
\nnlo{} (black, dotted) and \nll{}\plus{}\nlo{} (red, dashed) results are 
in reasonable agreement with the data, 
the \nnll{}\plus{}\nnlo{} result, however, being always closer to the data 
points in the low-\pt{} region where resummation effects are relevant; the 
\nlo{} central prediction (grey, dash-dotted), on the other hand, is quite off in that region.

\section{Conclusions and outlook}
\label{sec:summa}

We presented a general implementation of small-\pt{} resummation
in the {\sc Matrix} framework. Logarithmically enhanced contributions 
are resummed through \nnll{} accuracy and consistently combined with
the \nnlo{} cross section for any process with colorless final states, 
as long as the respective two-loop amplitude is known. 

In this proceedings article we further reviewed the first 
application of this framework to on-shell \ww{} and \zz{} production
\cite{Grazzini:2015wpa}, showing results for both the inclusive 
\pt{} distribution of the pair and within cuts on its momentum.
We also reported on results for the \pt{}-vetoed cross section 
and a comparison to experimental data of the \zz{} \pt{} spectrum.

Exploiting the helicity amplitudes of 
\citeres{Caola:2015ila,vonManteuffel:2015msa} to include the 
leptonic decays of the vector bosons with off-shell effects and 
spin correlations as well as the application to further processes
is left to future work.

\section*{Acknowledgement}
This research was supported in part by the 
Swiss National Science Foundation ({\abbrev SNF}) under contracts 
CRSII2-141847 and 200021-156585.

\renewcommand{\baselinestretch}{0}

\renewcommand{\em}{}
\bibliographystyle{polonica-mod}
\bibliography{Proceedings_Radcor}

\end{document}